\input phyzzx.tex
\tolerance=1000
\voffset=-0.0cm
\hoffset=0.7cm
\sequentialequations

\def\t1{{\tilde 1}}

\def\t{\theta}

\REF{\CAR}{J. L. Cardy,  Nucl. Phys. {\bf B270} (1986) 186.}
\REF{\VER}{G. J. Turiaci and H. L. Verlinde, JHEP {\bf 1612} (2016) 110 [arXiv:1603.03020].}
\REF{\STR}{M. Guica, T. Hartman, W. Song and A. Strominger, Phys.Rev. {\bf D80} (2009) 124008, [arXiv:0809.4266]; I. Bredberg,
C. Keeler, V. Lysov and A. Strominger, Nucl.Phys.Proc.Suppl. {\bf 216} (2011) 194, [arXiv:1103.2355].}
\REF{\CARL}{S. Carlip, Phys. Rev. Lett. {\bf 82} (1999) 2828, [arXiv:hep-th.9812013]; Class. Quant. Grav. {\bf 16} (1999) 3327,
[arXiv:gr-qc/9906126].}
\REF{\SOL}{S. Solodukhin, Phys. Lett. {\bf B454} (1999) 213, [arXiv:hep-th/9812056].}
\REF{\HAL}{E. Halyo, [arXiv:1502.01979]; [arXiv:1503.07808]; [arXiv;1809.10672].}
\REF{\POL}{A. Almheiri and J. Polchinski, JHEP {\bf 1511} (2015) 014, [arXiv:1402.6334].}
\REF{\STA}{J. Maldacena, D. Stanford and Z. Yang, PTEP 2016 (2016) no.12, 12C104, [arXiv:1606.01857].}
\REF{\CVE}{E. Verlinde, [arXiv:hep-th/0008140].}
\REF{\SYK}{A. Kitaev, http://online.kitp.ucsb.edu/online/entangled15/kitaev/, http:// \hfill
online.kitp.ucsb.edu/online/entangled15/kitaev2/,
J. Maldacena and D. Stanford, Phys.Rev. {\bf D94} (2016) no.10, 106002, [arXiv:1604.07818].} 
\REF{\ADS}{J. Maldacena, Adv. Theor. Math. Phys. {\bf 2} (1998) 231, [arXiv:hep-th/9711200]; S. Gubser, I. Klebanov and A. Polyakov, Phys. Lett. {\bf B428} (1998) 105, [arXiv:hep-th/9802109]; E. Witten, Adv. Theor. Math. Phys. {\bf 2} (1998) 253, [arXiv:hep-th/9802150].}
\REF{\BTZ}{A. Strominger, JHEP {\bf 9802} (1998) 009, [arXiv:hep-th/9712251].}
\REF{\SEN}{R. K. Gupta and A. Sen, JHEP {\bf 0904} (2009) 034, [arXiv:0806.0053].}
\REF{\VJY}{V. Balasubramanian, J. de Boer, M. M Sheikh-Jabbari and J. Simon, JHEP {\bf 1002} (2010) 017, [arXiv:0906.3272].}
\REF{\CAD}{M. Cadoni and S. Mingemi, Phys. Rev. {\bf D59} (1999) 081501, [arXiv:hep-th/9810251]; Nucl. Phys. {\bf B557} (1999)
165, [arXiv:hep-th/9902040].}



\singlespace
\pagenumber=0
\normalspace
\medskip
\bigskip
\titlestyle{\bf{The Cardy Formula from Goldstone Bosons}}
\smallskip
\author{ Edi Halyo{\footnote*{email: halyo@stanford.edu}}}
\smallskip
\centerline {Department of Physics} 
\centerline{Stanford University} 
\centerline {Stanford, CA 94305}
\smallskip
\vskip 2 cm
\titlestyle{\bf Abstract}
Two dimensional conformal field theories, can be described by their pseudo Goldstone bosons when the conformal symmetry is spontaneously and anomalously broken. We show that the Schwarzian action of these bosons leads to the Cardy formula without using modular invariance. As a result, the Cardy formula applies to conformal field theories on a cylinder and chiral theories in one dimension. This also explains why the Cardy--Verlinde formula for theories on $S^1 \times S^{d-2}$ can be written in the form of the Cardy formula of an effective two dimensional theory.

\singlespace
\vskip 0.5cm
\endpage
\normalspace

\centerline{\bf 1. Introduction}
\medskip

It is well--known that the entropy of a state in a 2D conformal field theory(CFT) is given by the Cardy formula[\CAR]. More precisely, the Cardy formula gives the degeneracy of a state with a fixed conformal weight when the CFT is on a spatial circle and at a finite temperature.
Since finite temperature corresponds to a compact Euclidean time direction, the CFT is considered to live on a Euclidean torus which
is invariant under modular transformations.
The modular symmetry of the CFT is used to obtain the partition function at high temperatures (which is hard to compute) directly from the simple one at zero temperature. The CFT entropy can be easily computed from this partition function and is given by the Cardy formula.

Recently it was realized that CFTs may be described by pseudo Goldstone bosons(PGBs) when the conformal symmetry is anomalously and spontaneously broken[\VER]. CFTs with nonvanishing central charges are anomalous. In addition, a state with nonvanishing conformal weight can be considered to be at finite temperature. This is due to the fact that a state with nonzero conformal weight has a large degeneracy and the CFT state is actually an ensemble of states. Following 
ref. [\VER], we consider a nonzero CFT temperature as a source of spontaneous breaking of the conformal symmetry.

As a result, we expect such a CFT to be described by the PGBs of conformal symmetry broken down to $SL(2)_R \times SL(2,R)_L$ where $SL(2,R)$ is the unbroken global conformal symmetry group (for each light--cone direction, $u$ and $v$). The PGBs realize the conformal symmetry nonlinearly and are parametrized simply by the chiral conformal transformations, i.e. $\xi(u),\eta(v)$. Their physics is described by (two copies of) the Schwarzian action which is $SL(2,R)_R \times SL(2,R)_L$ symmetric[\VER].
In this paper, we show that the Cardy formula can be derived from the Schwarzian action without resorting to the modular symmetry of the CFT. This means that the finite temperature CFT does not have to live on a torus.
Therefore, the Cardy formula applies to CFTs on a cylinder. Moreover, it also applies to 
CFTs in one compact dimension such as chiral CFTs and those on (compact) Euclidean time. These CFTs
play an important role in the description of different black holes[\STR-\STA].

These results also help to solve a puzzle related to the Cardy--Verlinde formula that gives the entropy of a CFT 
on $S^1 \times S^{d-2}$ whih is the boundary of $AdS_d$[\CVE]. It is known that this formula can be expressed as the Cardy formula for an effective CFT even though the higher dimensional CFT cannot be reduced to 2D. However, it can be reduced to 1D (by reducing on $S^{d-2}$) and our results indicate that the Cardy formula applies to the effective 1D CFT in the Euclidean time direction.

This paper is organized as follows. In the next section, we review the usual derivation of the Cardy formula. In section 3,
we derive the Cardy formula from the physics of the PGBs. In section 4, we explain, using our results, why the Cardy--Verlinde formula for higher dimensional CFTs can be written as the Cardy formula for 2D CFTs.
Section 5 contains a discussion of our result and our conclusions.

\bigskip
\centerline{\bf 2. The Cardy Formula for 2D CFTs}
\medskip

In this section, we review the derivation of the Cardy formula[\CAR] for the entropy of a (Euclidean) 2D CFT state. Consider a CFT 
with left and right central charges $c_L,c_R$ on a (Euclidean) torus defined by the identifications
$$(t_E,x) \sim (t_E+\beta,x+\theta) \sim (t_E,x+2 \pi) \qquad. \eqno(1)$$
Here $\beta=1/T$ and $2 \pi$ are the periodicities of the Euclidean time and space directions respectively and $\theta$ is the twist.
In terms of the complex coordinate $z=(it_E+x)/2 \pi$ the identifications in eq. (1) become $z \sim z+1$ and $z \sim z+ \tau$ where we defined the modular parameter of the torus $\tau=(i \beta+\theta)/2 \pi$.

A CFT state is defined by its left and right conformal weights $L_0$, ${\bar L_0}$. It is well--known
that if the CFT is on a circle, due to the Casimir effect, the energy of the states or the conformal weights are shifted to $L_0^{\prime}=L_0-c_R/24$ and
${\bar L_0}^{\prime}={\bar L_0}-c_L/24$. The partition function of this CFT is
$$Z= Tr e^{2 \pi i \tau L_0^{\prime} -2 \pi i {\bar \tau} {\bar L_0}^{\prime}}= Tr e^{-\beta H+ i \theta J} \qquad, \eqno(2)$$
in terms of the Hamiltonian $H=L_0^{\prime}+{\bar L_0}^{\prime}$ and spin $J=L_0^{\prime}-{\bar L_0}^{\prime}$. In the following we set $\theta=0$ for simplicity. The derivation of the Cardy formula below can be easily generalized to
$\theta \not =0$.

We now concentrate on the right moving sector of the CFT since the treatment of the left moving sector is identical. The partition function for the right movers can be written as 
$$Z=Tr e^{-\beta L_0^{\prime}} \qquad. \eqno(3)$$
At low temperatures, $T \to 0$, only the vacuum state with $L_0^{\prime}=-c/24$ survives in the trace over states and thus
$$Z=e^{\beta c/24} \qquad. \eqno(4)$$
In order to obtain the partition function at high temperatures, we use the modular symmetry of a CFT on a torus. The torus on which the CFTs lives (see eq. (1)) is invariant under the $SL(2,Z)$ transformations of the modular parameter defined by (with integer $a,b,c,d$)
$$\tau \to {{a \tau+ b} \over {c \tau+ d}} \qquad ad-bc=1 \qquad. \eqno(5)$$
Therefore, the CFT partition function in eq. (2) is also modular invariant. 

It is hard to calculate the high temperature partition function with $\beta \to 0$ directly from eq. (3) due to large number of states that contribute to it. However, this can be easily done by using the modular symmetry of the CFT. 
In our case, $\theta=0$ and thus one of the generators of the transformation in eq. (5), $S: \tau \to -1/\tau$ 
becomes $\beta \to 4 \pi^2/\beta$ which is a high/low temperature duality. Then, using eq. (4), we find that
$$Z(\beta)=Z(4 \pi^2/ \beta)=e^{\pi^2 c/6 \beta} \qquad. \eqno(6)$$
We can easily obtain the energy and entropy from this partition function 
$$E=-\partial_{\beta} logZ={{\pi^2 c} \over {6 \beta^2}} \qquad, \eqno(7)$$
and
$$S=(1-\beta \partial_{\beta}) log Z={{\pi^2 c} \over {3 \beta}} \qquad. \eqno(8)$$
Alternatively, we can write the partition function as (where $E=L_0^{\prime}$)
$$Z(\beta)=\int dE \rho(E) e^{-\beta E} \qquad, \eqno(9)$$
where $\rho(E)=e^{S(E)}$ is the density of states. A Laplace transform gives
$$\rho(E)=\int d\beta e^{\beta E} Z(\beta) \qquad. \eqno(10)$$
Using eq. (6) and the saddle point approximation we get
$$T={1 \over {2 \pi}} \sqrt{{{24 L_0^{\prime}} \over c}} \qquad, \eqno(11)$$
and
$$S(E)=log \rho(E)= 2\pi \sqrt{{c\over 6} \left(L_0-{c \over {24}} \right)} \qquad, \eqno(12)$$
where we used $E=L_0^{\prime}=L_0-c/24$.

The above derivation was only for the right moving sector. Taking into account the identical contribution of the left moving sector we get from eqs. (8) and (12) the two forms of the Cardy formula for the total entropy 
$$S={{\pi^2 c_R} \over {3 \beta_R}}+{{\pi^2 c_L} \over {3 \beta_L}} \qquad, \eqno(13)$$
or
$$S=2\pi \sqrt{{c_R\over 6} \left(L_0-{c_R \over {24}} \right)}+2\pi \sqrt{{c_L\over 6} \left({\bar L_0}-{c_L \over {24}} \right)} \qquad. \eqno(14)$$
These two Cardy formulas are valid for high temperatures $\beta<<1$ or as can be seen from eq. (11) for $L_0>>c$. Eq. (13) describes the entropy of a 2D gas (in the canonical ensemble) whereas eq. (14) describes the different number of ways a state with $L_0,{\bar L_0}$ can be built from raising operators (in the microcanonical ensemble).


\bigskip
\centerline{\bf 3. The Cardy Formula from the Pseudo Goldstone Bosons of CFT} 
\medskip

The derivation of the Cardy fomula above was based on the high energy partition function which was obtained using modular symmetry of the CFT. In this section, we obtain the same partition function from the physics of the pseudo Goldstone bosons of conformal symmetry described by their Schwarzian action. As in the previous section, we consider only the right moving part since the treatment of the left moving part is identical.

Any CFT state with $L_0 \not=0$ can be essentially considered as an ensemble of states at a finite temperature (given by eq. (11)) which we consider to be the spontaneous breaking of conformal symmetry. 
In addition, as noted above, if the central charge does not vanish, the CFT is anomalous.
The anomaly shows up in the commutators of the Virasoro generators
$$[L_n,L_m]=(n-m)L_{n+m}+{c \over {12}} n(n^2-1) \delta_{n+m,0} \quad, \eqno(15)$$
where the second term proportional to the central charge is the anomaly. From eq. (15) we see that there is no anomaly for 
$L_0, L_{\pm 1}$.
Thus, the full conformal symmetry is anomalously broken down to the global conformal symmetry or $SL(2,R)$. As a result of the spontaneous and anomalous
breaking of the conformal symmetry, at low energies, physics can be described by the PGBs. Since the global conformal symmetry remains unbroken, we expect the PGB Lagrangian to be $SL(2,R)$ invariant. Moreover, we expect the kinetic and interaction terms for the PGBs to be proportional to the spontaneous and anomalous symmetry breaking parameters $T$ and $c$ respectively.

In ref. [\VER], it was shown that the PGBs of conformal symmetry can be described exactly by the reparametrizations 
which realize the conformal transformations nonlinearly, i.e. $t_E-x=u \to \xi(u)$. Note that $u$ is periodic, $u \sim u+ \beta$.
The spatial direction $x$ may or may not be compact as in eq. (1) as long as $\beta<<2 \pi$, i.e. the CFT is at a high temperature.
Then, the energy--momentum tensor for $\xi(u)$ is given by
$$T(\xi)=L_0^{\prime} \xi^{\prime 2}+{c \over {12}} Sch(\xi,u) \quad, \eqno(16)$$
where the Schwarzian derivative is 
$$Sch(\xi,u)=\left({\xi^{\prime \prime} \over {\xi^{\prime}}} \right)^{\prime}- {1 \over 2} {\xi^{\prime \prime 2} \over {\xi^{{\prime}2}}} \quad, \eqno(17)$$
and prime denotes a derivative with respect to $u$. We see that, as expected, the kinetic terms is proportional to 
$L_0^{\prime} \sim T^2$ and the Schwarzian interaction term appears with the central charge $c$ in the action. 

Eq. (16) for the energy leads to the action[\POL,\STA]
$$I_{\xi}=- \int_0^{\beta} du \left ({L_0^{\prime}}\xi^{\prime 2}+ {c \over {12}}Sch(\xi,u) \right) \quad, \eqno(18)$$
with the equation of motion
$$2L_0^{\prime} \xi^{\prime \prime}+{c \over {12}}{(Sch(\xi,u))^{\prime} \over {\xi^{\prime}}}=0 \qquad. \eqno(19)$$
The solution is $\xi(u)= \alpha u$ (up to an irrelevant additive constant) where $\alpha$ is a constant. Note that the action in eq. (18) is not $SL(2,R)$ symmetric due to the 
kinetic term. However, it is easy to see that, any rescaling of a solution to the equation of motion is also a solution. We can use this rescaling freedom to normalize the kinetic term so that the action now becomes
$$I_{\xi}=- {c \over {12}} \int_0^{\beta} du \left ({1 \over 2}\xi^{\prime 2}+ Sch(\xi,u) \right) \quad. \eqno(20)$$
This action is still not $SL(2,R)$ symmetric but can be put in a symmetric form in terms of new PGB fields defined by $\phi=tan(\xi/2)$ with the action[\STA]
$$I_{\phi}=- {c \over {12}} \int_0^{\beta} du Sch(\phi,u)  \quad. \eqno(21)$$
This action is $SL(2,R)$ symmetric. Thus, it is really the $\phi$,  rather than $\xi$, that are the true PGBs of the CFT. $\phi$ parametrizes the space $Virasoro/SL(2,R)$, i.e. the Virasoro group modulo the Mobius group.
The equation of motion obtained from the action in eq. (20) is the same as that in eq. (19) where $L_0^{\prime}$ is replaced by $c/24$. The solutions to both equations are the same, i.e. $\xi(u)=\alpha u$.

In order to determine the constant $\alpha$ we demand that the energy of the PGB solution match the conformal weight $L_0^{\prime}$
(or the energy of the CFT in eq. (7))
$$E_{\xi}={c \over {12}} \left ({1 \over 2}\xi^{\prime 2}+ Sch(\xi,u) \right)=L_0^{\prime} \quad. \eqno(22)$$
We find that $\alpha=\sqrt{24 L_0^{\prime}/c}=2 \pi T=2 \pi /\beta$. 
Substituting the solution $\xi(u)=(2 \pi /\beta)u$ into eq. (20) we find
$$-I_{\xi}= log Z(\beta)= {{\pi^2 c} \over 6{ \beta}} \qquad, \eqno(23)$$
which agrees with eq. (6). This is our main result: the high temperature CFT partition function can be obtained from the Schwarzian action for the PGBs. The left moving sector of the CFT gives an identical partition function and the derivation of the two versions of the Cardy formula in eqs. (13) and (14) is identical to the one at the end of section 2. Thus, the 2D gas with the entropy given by eq. (13) is actually a gas of PGBs. The gas is strongly interacting as can be seen from the Schwarzian action. Nevertheless, it acts like a free gas with a fixed number of degrees of freedom given by the central charge.

Note that we derived the Cardy formula from the Schwarzian action without using modular invariance which means that the CFT does not have to live on a torus. Thus, the Cardy formula also applies to CFTs that live on a cylinder or only in one compact dimension such as chiral CFTs and those on Euclidean time. Even though the CFT is one dimensional, the entropy in eq. (13) has the form of that of a 2D gas. The 1D CFT somehow has the memory of its 2D origin.

Above, we rescaled $\xi$ in order to match the coefficients of the kinetic and Schwarzian terms in the action.
If we did not rescale the PGBs $\xi$ as in eq. (20), the energy condition in eq. (22) would give $\alpha =1$ leading to the partition function $log Z= L_0^{\prime}  \beta$. Using eq. (8) this seems to imply a vanishing entropy. However, using eq. (11) for $\beta$
we obtain the correct partition function in eq. (23). In fact, we could use any rescaling of the PGBs and obtain the correct partition function as long as we satisfy the energy condition in eq. (22).

Any one dimensional conformal theory is effectively described by the Schwarzian action since this is the simplest action invariant under the unbroken symmetry, i.e. $SL(2,R)$. The conformal symmetry does not have to be fundamental as it is in CFTs. For example, it may arise in the low energy limit as in the SYK model[\SYK] or it may be a geometrical symmetry as for the
near $AdS_2$ space--times[\STA]. Our results imply that the coefficient of the Schwarzian derivative in the action 
should be set equal to $-c/12$ in addition to the identification in eq. (22). For the Schwarzian action describing near $AdS_2$ spaces this coefficient is $-\phi_b/8 \pi G_2$ where $\phi_b$ is the boundary value of the dilaton[\STA]. On the other hand, for $AdS_2$ the central charge of the asymptotic (chiral) conformal symmetry is $c/12=1/8 \pi G_2$. We see that for $\phi_b=1$ the coefficient of the Schwarzian action is exactly $-c/12$ as required. For the SYK model, the coefficient is (up to numerical factors of O(1)) 
$\sim N/J$ where $N$ is the number of Majorana fermions and $J$ is the interaction strength[\SYK]. Thus, in this case we have
$c/12 \sim N/J$. For $N \geq 4$ as in the SYK model, we find that the central charge is inversely proportional to $J$.




\bigskip
\centerline{\bf 4. The Cardy-Verlinde Formula as the Cardy Formula}
\medskip

The Cardy formula for the entropy of a 2D CFT can be generalized to a CFT living on $S^1 \times S^{d-2}$ by using the AdS--CFT correspondence[\ADS]. In this section, we argue that this Cardy--Verlinde formula[\CVE] can be understood as the entropy arising from the Schwarzian action given by eq. (20).

Starting from the metric of a black hole in $AdS_d$ one can derive the Cardy--Verlinde formula for the entropy in the dual finite temperature CFT that lives on the boundary $S^{d-2}$[\CVE] 
$$S_{CFT}={{4 \pi} \over {(d-2)}} R \sqrt{E_E E_C} \qquad, \eqno(24)$$
where the extensive and Casimir energies, $E_E$ and $E_C$ respectively, add up to the total CFT energy as
$$E_{CFT}=E_E+E_C={{c(d-2)} \over {48 \pi}} {V \over {L^{d-1}}} \left(1+ {{L^2} \over {R^2}} \right) \qquad, \eqno(25)$$
where $V=A_{d-2} R^{d-2}$ is the volume of the boundary sphere (which has been rescaled to match the AdS black hole area) with $A_{d-2}$ the unit volume in $d-2$ dimensions,
In eq. (25), $R$ and $L$ are the black hole and AdS radii respectively and $c=3L^{d-1}/G_d$.

Eq. (24) for the entropy can be written in a more suggestive form by defining a central charge, $C$, for an effective CFT and a conformal weight $L_0^{\prime}$ for a CFT state
$${C \over {24}}={{RE_C} \over {(d-2)}}  \qquad  L_0^{\prime}={{RE_E} \over {(d-2)}} \qquad, \eqno(26)$$
where $L_0^{\prime}=L_0-C/24$ and $L_0=E_{CFT}$.
Eq. (24) can now be written as
$$S_{CFT}=2\pi \sqrt{{C \over 6} \left(L_0-{C \over {24}} \right)} \qquad, \eqno(27)$$
which is precisely the Cardy formula for a 2D CFT.

It has been difficult to understand this form of the CFT entropy since the boundary CFT lives on $S^1 \times S^{d-2}$ and not in 2D. It does not appear possible to reduce this CFT to 2D either.
However, we can reduce the higher dimensional CFT on $S^{d-2}$ and obtain a CFT that lives in the 
the periodic Euclidean time direction. In section 2, we saw that such a CFT
is described by PGBs with the Schwazian action. The entropy arising from this action is exactly the Cardy formula.
The effective CFT is simply the chiral CFT obtained by reducing on $S^{d-2}$. 
Thus, the appearance of an effective 2D Cardy formula for higher dimensional CFTs is due to the fact that the Cardy formula 
works for CFTs on compact Euclidean time. 

The simplest example of the above result, which is also best understood case, is the relation between the 2D nonchiral CFT that lives on the boundary of $AdS_3$ and the chiral 1D CFT that lives on the boundary of $AdS_2$[\BTZ,\SEN,\VJY,\CAD]. The BTZ black hole can be described by a nonchiral CFT that lives on the
2D boundary of $AdS_3$. In this case, the Cardy--Verlinde formula reduces to the Cardy formula with 
$C=3L^2/G_3=c$ and $L_0^{\prime}=R M_{BH}$.
The boundary CFT can be compactified on the boundary $S^1$ giving rise to a 1D CFT that lives on
the boundary of $AdS_2$ which is the compact Euclidean time. The BTZ black hole becomes an $AdS_2$ black hole with the same entropy.
As we saw in section 2, this theory is described by the Schwarzian action and leads to the Cardy formula. This case is somewhat trivial since the``higher dimensional" CFT is two--dimensional and it is not surprising that the Cardy formula works. However, the appearance of the chiral 1D CFT with the same entropy is a nontrivial check on the ideas in the previous paragraph.
The BTZ black hole (or $AdS_3$ case) is the simplest example but it seems that the same argument applies to CFTs in higher dimensions. It would be nice to show this result explicitly even though we do not understand the CFTs well enough except for the $D=3,4$ cases.


\bigskip
\centerline{\bf 5. Conclusions and Discussion}
\medskip

In this paper, we showed that the thermodynamics of 2D CFTs (with an anomaly and at a finite temperature) can be described by their PGBs with a Schwarzian action. This can be seen as further supporting evidence for the description of CFTs in terms of their PGBs.
The CFT entropy given by eq. (13) is that of a 2D gas of PGBs. The derivation of the Cardy formula using the Schwarzian action does not require the modular invariance of the CFT on a torus. Therefore, the Cardy formula applies to CFTs on a cylinder. In this case, all our formulas remain the same with the replacement of $L_0^{\prime}$ by $L_0$ since there is no shift in the conformal weights due to the Casimir effect. 

In addition, since modular invariance is not necessary, the Cardy formula 
also applies to CFTs that are defined in one (compact) dimension, e.g. chiral CFTs on a circle or those in a compact Euclidean time direction. These types of CFTs play an important role in the description of different black holes. For example, extremal Kerr[\STR] and generic
nonextremal black holes[\CARL-\HAL], including the Schwarzschild black hole, can be described by chiral CFTs and CFTs on Euclidean time respectively. Near extremal Reissnner--Nordstrom
black holes are described by the Schwarzian action[\POL,\STA] which, as we saw above, is a low energy description of chiral CFTs.
The entropies of all these black holes are reproduced by the Cardy formula even though they strictly live in one dimension. 

We also showed that the description of CFTs in one compact dimension in terms of the PGBs resolves a puzzle related to the 
Cardy--Verlinde formula for CFTs on $S^1 \times S^{d-2}$ which are dual to $AdS_d$ black holes. The Cardy--Verlinde formula can be written as the Cardy formula for an effective CFT even though $S^1 \times S^{d-2}$ cannot be reduced to two dimensions. 
However, we can reduce the theory on $S^{d-2}$ and obtain a 1D CFT 
living on the compact Euclidean time direction (remaining on the boundary). As we saw above, this is
described by the Schwarzian action for the PGBs and with an entropy given by the 2D Cardy formula.

We used the Schwarzian action for the PGBs to describe the thermodynamics of CFTs. It is puzzling that we obtained the high temperature partition function from the Schwarzian action for the PGBs that describes the CFT at low energies.
PGBs should describe the physics around $L_0^{\prime} \sim 0$ and not in the thermodynamic limit for which $L_0>>c$. 
Our results indicate that perhaps we should consider, rather than the whole CFT, each level of the CFT with $L_0^{\prime} \not =0$ as a separate theory. In that case, each level behaves like a different theory in which the conformal invariance is spontaneously and anomolously broken. PGBs then describe the large number of states at or slightly above a given $L_0^{\prime}$ which is the vacuum of the level.


\bigskip
\centerline{\bf Acknowledgments}

I would like to thank the Stanford Institute for Theoretical Physics for hospitality.

\vfill

\refout

\end
\bye